\documentclass[a4paper,11pt,fleqn]{article}

\usepackage[utf8]{inputenc}
\usepackage[T1]{fontenc}
\usepackage{textcomp}
\usepackage{amsmath,amssymb,amsthm,mathrsfs}
\usepackage{authblk}
\usepackage{lmodern}
\usepackage[a4paper]{geometry}
\usepackage{graphicx}
\usepackage[usenames,dvipsnames]{xcolor}
\usepackage{microtype}
\usepackage{enumerate}
\usepackage{hyperref}
\usepackage{authblk}
\usepackage{algorithm2e}
\usepackage{pgf,tikz}
\usepackage{float}
\usepackage{latexsym,amsmath,amsfonts,amscd,amssymb}
\usepackage{graphicx}
\usepackage[font=footnotesize,labelfont=bf]{caption}
\hypersetup{pdfstartview=XYZ}
\newtheoremstyle{theoreme}
 {\topsep} 
 {\topsep} 
 {} 
 {0pt} 
 {\bfseries} 
 {\newline} 
 {3pt} 
 {} 
\newtheoremstyle{proposition}
{\topsep} 
 {\topsep} 
 {} 
 {0pt} 
 {\bfseries} 
 {\newline} 
 {3pt} 
 {} 
 \newtheoremstyle{lemma}
 {\topsep} 
 {\topsep} 
 {} 
 {0pt} 
 {\bfseries} 
 {\newline} 
 {3pt} 
 {} 
\newtheoremstyle{definition}
 {\topsep}
 {\topsep}
 {}
 {0pt}
 {\bfseries}
 {\newline}
 {3pt}
 {}
\newtheoremstyle{remarque}
 {\topsep}
 {\topsep}
 {}
 {0pt}
 {\bfseries}
 {: }
 {3pt}
 {} 
\theoremstyle{theoreme}
\newtheorem{Thm}{Theorem}[section]
\newtheorem*{Thm*}{Theorem}

\theoremstyle{lemma}

\theoremstyle{proposition}

\newtheorem*{Prop*}{Proposition}

\theoremstyle{definition}

\theoremstyle{remarque}
\newtheorem{Rem}[Thm]{Remark}

\newtheorem*{Claim*}{Claim}

\newtheorem*{Remark*}{Remark}

\newtheorem*{Application*}{Application}

\title{Ant Routing scalability for the Lightning Network}
\author[,1]{Cyril Grunspan, Gabriel Lehéricy
\thanks{Emails: \texttt{cyril.grunspan@devinci.fr},\quad
\texttt{gabriel.lehericy@devinci.fr},\quad}}
\author[,2]{Ricardo Pérez-Marco
\thanks{Email: \texttt{ricardo.perez.marco@gmail.com}}}
\affil[1]{De Vinci Research Center, Paris-La Défense, France}
\affil[2]{CNRS, IMJ-PRG, Universit\'e Sorbonne, Paris, France}
\begin{document}
\providecommand{\keywords}[1]{
\smallskip
\noindent
\textbf{Keywords.} #1}
 \maketitle
 
 \abstract{The ambition of the Lightning Network is to provide a second layer to the Bitcoin network to enable 
 transactions confirmed instantly, securely and anonymously with a world scale capacity using a decentralized 
 protocol. Some of the current propositions and implementations present some difficulties in anonymity, scaling and decentalization. The Ant Routing algorithm for the Lightning Network was proposed in \cite{GrunspanPerez} for maximal decentralization, anonymity 
 and potential scaling. It solves several problems of current implementation, such as channel information update and centralization by beacon nodes.
 Ant Routing nodes play all the same role and don't require any extra information on 
 the network topology beside for their immediate neighbors. The goal of LN transactions are completed instantaneously 
 and anonymously. We study the scaling of the Ant Routing protocol. We propose a precise 
 implementation, with efficient memory management using AVL trees. 
 We evaluate the efficiency of the algorithm and we estimate the memory usage of nodes by local 
 node workload simulations. We prove that the number of transactions per second that Ant Routing can sustain is of the order of several thousands which is enough for a global payment network.}
 
 \section{Introduction}

 The Lightning Network was introduced in order to address Bitcoin scalability \cite{LNwhitepaper}. It consist on 
 a ``layer 2''-network set on top of Bitcoin's network which facilitates micro-payments between bitcoin users without needing
 to broadcast the transaction on the blockchain.
 
 The Lightning Network is not a complete graph, but has the property of channel composition. This allows any user on the network 
 to make secure payments to any other node in the same connected component of the network, even when no direct payment channel exists.
 For example, assume that Alice and Bob are both LN users, but have no direct payment channel between them. 
 Now assume that there is a third user, Charlie, who has a direct channel with Alice and a direct channel with Bob. Then 
 Alice can pay to Bob via Charlie: Alice sends the payment to Charlie, who will then forward it to Bob. All these payments are realized in 
 a secured way by the protocol of channel composition. 
 This allows any LN node to send money to any other LN node when the network is connected.
 
 Obviously this requires finding a payment route  from Alice to Bob, hence the need for a good routing algorithm. 
 Ideally, one wants to have a routing algorithm which is safe, decentralized, anonymous, and only requires minimum knowledge 
 of the network from all participants. In \cite{GrunspanPerez}, the authors proposed a new routing algorithm, called \textit{Ant Routing}.  Ant Routing is a completely decentralized routing algorithm which does not require beacon nodes nor routing tables. Instead, each node of the network executes the exact same task which requires no knowledge of the topology of the network.
 Moreover, it is worth noting that using Ant Routing on the lightning network would have a positive impact on  fees. Indeed, unlike the current routing mechanism in use, ant routing does not allow for a node to ask for fees proportional to the transaction amount.  This also strengthens the network's decentralization.

 The goal of this article is to estimate the performance of Ant Routing, and  
 propose more precise and efficient implementation. In particular, we investigate if this routing 
 algorithm is compatible with the original ambition of the Lightning Network to solve Bitcoin scalability at planetary level, hence 
 providing an instantaneous decentralized global payment network. 
 In practice, the question is whether Ant Routing algorithm can allow the network to sustain several thousands of transactions 
 per second. This article gives a positive answer to this question. The conservative estimates presented in Sections \ref{proxymodelsection} and \ref{experimentalsection} show that  Ant Routing can  sustain more than $10.000$ transactions per second.

     The paper is organized as follows. In Section \ref{Overviewsection}, we give 
     an overview of routing protocols which have been proposed for the Lightning Network and of algorithms based on ant behavior.     
     Section \ref{protocolsection} explains the ant routing protocol. We start by briefly recalling the rough ideas of ant routing 
     as presented in \cite{GrunspanPerez}. We then give a more detailed description of the Ant Routing protocol. In particular,  we 
     give a precise list of data that must be kept in memory and a list of tasks performed by nodes. 
     In Section \ref{lifetimesection} we estimate the duration for seeds to be kept in memory. 
     In section \ref{proxymodelsection} we give a general argument to determine network capacity for networks with a 
     similar structure than Bitcoin. This argument applies to Ant Routing provided that 
     the local process time for transactions 
     in the nodes is neglectable or of the same order of magnitude than the propagation time across the network. We propose efficient local algorithms and run simulations.  
     In Section \ref{AVLsection}, we review AVL trees, which is the data structure used for the  Ant Routing implementation. In Section \ref{Implementationsection}, we propose a concrete implementation of Ant Routing. Finally, in Section \ref{experimentalsection} we estimate the computing time and memory usage. Putting together these results, we conclude by an estimate on the number of transactions per second that the proposed implementation of Ant Routing can sustain.

     \section{Overview of routing proposals and existing ant algorithms}\label{Overviewsection}
     
        The white paper for the Lightning Network is sketchy about routing and no  specific routing protocol is proposed \cite{LNwhitepaper}. 
        The authors believe that routing tables are necessary, and in fact, it seems every routing protocol 
        proposed for the Lightning Network prior to \cite{GrunspanPerez} makes use of routing tables.
        We review in this section some of these proposals. 
        A review of routing solutions for the lightning network can also be found in \cite{Kohalli}.        
        One should keep in mind that, in accordance to Bitcoin philosophy, one wishes 
        the Lightning Network to remain decentralized and to preserve confidentiality of payments.
        
         One major constraint for routing in the Lightning Network is to find a path with sufficient capacity. It seems natural to try to solve this problem via 
         a distributed max-flow approach which computes the 
         path with the largest capacity
         \cite{DistributedMaxFlow,Maxflow,Maxflow2}. However,          
         max-flow routing performs poorly for large network, as its run time growths quadratically  on the number of channels 
         \cite{Maxflow2}. Another possibility is to use 
        \textit{beacons} instead.          
        A beacon is a special node whose role is to enhance the other nodes' visibility of the network and help them in 
        route computation.  One example of an algorithm which uses beacons is \textit{Flare} \cite{Flare}.
        Another classical approach to routing is 
        \textit{landmark routing}.  Landmark routing has been adapted to 
        payment network in algorhitms such as \textit{SilentWhispers} \cite{SilentWhisper}.
        The idea of landmark routing is to compute a route from the 
        source to the destination via a previously chosen node, called a \textit{landmark}, 
        which is usually  a highly connected node.
        The main drawback 
        of having a system rely on landmarks or beacons is the potential threat to decentralization, as well as its fragility to attacks.
        
        Another approach to routing is  \textit{embedding-based routing}. The idea is to embed nodes into a vector space such that 
        the hop distance between nodes 
        is reflected by their distance in the vector space. For routing, the source node chooses the neighbor whose vector 
        representation is the closest to the vector representation of the destination node. Each node in the path then repeats this 
        strategy. This requires regularly updating information about the network. An example of such routing algorithm is \textit{Speedy Murmurs}, which is based on \textit{Voute} \cite{Voute,SpeedyMurmurs}.

       Recent attention has been given to payment routing solutions that keep 
       channels balanced. Keeping channels balanced is considered important  in the current conception of the 
       Lightning Network\footnote{A network with only one directional channels is theoretically more limited, 
       but in practical terms can be useful since 
       most of the channels are used unidirectionnally.}: if a channel 
       becomes too unbalanced, then payment can only flow in one direction, 
       and this restricts  payment routes that can no longer use that channel. In \cite{Spider},  
       the authors introduce an interesting idea  which considers channels' balance for route seclection called \textit{Spider}. 
       More precisely, \textit{Spider} splits payments into several smaller payments, each of which can be sent 
       through a different route. The route for each of these payment is chosen to optimize the re-balancing of unbalanced channels.
       
       Note that all of these routing strategies either require giving some nodes a special role (as a landmark or a beacon) or require each node to keep some 
       knowledge of the network topology. This knowledge is a vector of attack for the Lightning Network. One of the reasons for the proposal of
       Ant Routing is to obfuscate the topology of the network: It requires no landmark and requires no knowledge 
       on the topology, while at the same time preserves anonymity and decentralization. The decentralization of the 
       algorithm is achieved by making every node play exactly the same role in the routing process and using only knowledge about its neighbors. 
       
       The algorithm is inspired by the behavior of ants.
       Ant colonies have raised the interested of ethologists by their ability to optimize their route from nest to food source. Although 
       each ant individually seems to follow a random motion, their collective behavior finds efficiently the shortest path 
    from their nest to a food source. This is achieved through a ``\textit{stygmergic}'' communication of the ants with their environment through 
    pheromones \cite{Dorigo1}. More precisely, each walking ant leaves a trail of pheromone, which has the 
    attracts other ants to its path.
    
    The author of \cite{Dorigo1} and \cite{Dorigo2} describes the following experiment to explain this remarkable observed ant optimization behavior. We study an ant colony connected to a food source by two paths, a short  
    and a longer one. At the outset of the experiment, each ant walks in a random direction, choosing each path randomly 
    with equal probability. As they walk, the ants leave a pheromone trail behind.
    Because ants will travel faster on the shorter path than on the longer one, pheromones accumulate faster on the shorter path. 
    The concentration of pheromone on the shorter path increases, so does the probability that ants to choose this path. This leads to a 
    positive feedback, and eventually few ants will choose the longer path.
    
    The author of \cite{Dorigo1} had the idea to simulate this ant behavior with a multi-agent stochastic learning model 
    to solve optimization problems. An application to network routing problems is given in chapter 6 of \cite{Dorigo1}.
    
    The Ant Routing algorithm is inspired from these ant behavior models, but there are some major differences 
    with the routing propositions in \cite{Dorigo1} and \cite{Dorigo2}. Our algorithm is simpler and deterministic, and not probabilistic. 
    It only uses the idea of leaving pheromones to trace a path, but does not consider accumulation of pheromones on a path. 
    Eventually, the model could be enhanced taking into account pheromone accumulation to improve the efficiency.
    The paths  are not chosen randomly. The pheromones are flooding the network. 
    Also, apparently the applications of ant algorithms to routing problems in \cite{Dorigo1} and \cite{Dorigo2}  use  routing tables, which we want to avoid. 
 
 \section{Ant routing protocol}\label{protocolsection}
 
 We briefly recall the main ideas of ant routing as presented in \cite{GrunspanPerez}. We assume that Alice and Bob 
 both have a node on the Lightning Network and they have a direct communication channel but not necessarily a payment channel. 
 Alice wants to pay Bob via the Lightning Network, so they need to find a payment route in the Lightning Network. 
 
 Alice and Bob start by agreeing on a random number $S$, chosen big 
     enough to avoid collisions. Alice creates the number $S(0)=0^{\frown}S$ and Bob creates the number 
     $S(1)=1^{\frown}S$. $S(0)$ and $S(1)$ are called \textit{pheromone seeds}. Alice and Bob send their respective seeds
     to the network through their neighbors. Pheromone seeds propagate on the network being successively 
     forwarded by all nodes to their neighbors. When $S(0)$ and $S(1)$ meet at some node we have a match. 
     Each pheromone seed is forwarded with a counter associated. This counter is increased by $1$ at each hop. 
     If a node receives the same pheromone seed several times, then only the pheromone seed with lower 
     counter is stored and forwarded. The nodes keep track from whom they received the stored seeds. 
     When we have a match, the node where it occurs creates matched seeds $M(0)$ and $M(1)$, and sents back 
     to the respective neighbors from whom he received $S(0)$ and $S(1)$. Nodes receiving a matched seed forward them 
     in the same way, and after a finite number of steps, bounded by the counters, the matched seeds reach Alice and Bob. 
     
     After some time, Alice will receive several matched seeds, each corresponding to a payment route. 
     Alice is free to choose any of these matched seed as a payment
route. Note that, in order to preserve anonimity, it is desirable for Alice to choose
a route with at least two intermediary nodes.
     She selects one, by a selection algorithm that she is free to decide. Creates a confirmed seed $C$ and sends $C$ 
     back to her neighbor from whom she received the matched seed. Nodes forward confirmed seed they receive as for matched seeds. 
     Hence, the confirmed seed will reach Bob though a unique path. Once Bob has received $C$, he informs Alice and she 
     can proceed with the payment. The payment is then done via this path. 
     
     Notice that this ``path discovering algorithm'' finds a connecting path without anyone knowing this path. The only knowledge 
     the participating nodes have is about their neighbors, and they don't even know if these are Alice or Bob. 
     
     During the pheromone phase, a node may be tempted to cheat with the counter in order to make the path more 
     attractive. For this reason, we require a verification phase  after the confirmation phase and before the payment. 
     The point of this verification phase is to check that no node on the path indicated by $C$ has cheated with the 
     counter. This can be done in the following way: during the confirmation phase, each node, starting from Alice neighbor, 
     is asked to generate a random number. 
     This random number is appended to a list $l$, which lists all the random numbers from previous nodes. Then the nodes 
     forwards $l$ along with $C$ to the next node. When $C$ and $l$ reach Bob, Bob sends $l$ back and communicates the list 
     of random numbers to Alice. The nodes forward it back only if their random number is in the list. Then $l$ can only reach 
     back Alice through the nodes in the path if nobody has cheated on the counter.

   \medskip

    We make precise the task done by the nodes in the protocol. 
 It is important to note that all nodes, apart from Alice and Bob in their payment, execute the same task.
 Moreover, the nodes have no knowledge about the topology of the network. The only topological information a node needs is who are its direct neighbors.
 
 We list in the next section  which data the nodes must keep in memory, and which must be relayed (Section \ref{memorysubsection}). 
 We then give the algorithms that the nodes perform in each  phase of the ant routing protocol: 
 the pheromone phase, the match phase, the confirmation phase and the counter check phase (Section \ref{algorithmsection}).

 \subsection{Memory data}\label{memorysubsection}
 
   Each node allocates some memory space for the routing task. 
   This memory space contains three trees. The first one is used to store 
   pheromone seeds, the second one for matched seeds, and the third one for confirmation seeds.
   When she starts her routing task, Alice creates a fourth tree in her memory, which we call the 
   \textbf{special match tree}. This tree will be destroyed once Alice 
   has completed her payment. This tree is used to 
   store the matched seeds created from her own pheromone seed. These matches are therefore stored separately
   from other matched seeds.
   
     Each seed comes with some data. More precisely, for 
  each pheromone seed $P$, a node stores: 
 \begin{enumerate}[(1)]
 \item The pheromone seed $P$.
  \item A counter $c$.
  
  \item The id $s$ of the node from which it received the seed .
  \item The remaining fees $f$: at the start, Alice chooses a maximal amount $f_{max}$ of fees which she is willing to pay for the transaction. If $g$ denotes   
   the sum of all the fees of all nodes through which the seed has traveled, then the remaining fees $f$ is 
   $f=f_{max}-g$.
   \item The amount $a$ of the transaction.
 \end{enumerate}
 
  All this information is required for the functioning of the ant routing algorithm. The id $s$ is needed for the ``match'' phase, 
  when the matched seed will have to trace the way back to Alice. 
  The amount of the transaction is used by the relaying nodes to select channels with sufficient funds.
  The remaining fees are used by Alice at the start of the confirmation phase in her choice of the matched seed.
    Finally, the counter is used to tame the flood of data on the network (see section \ref{algorithmsection}).
 
 The node keeps $s$  to itself to preserve anonymity. The nodes also relay one extra piece of information 
 which is not stored in memory: a timestamp  $t$, which is decided upon by Alice at the creation of the seed. 
 This timestamp allows the node to know the age of the seed, which is necessary information so that the the node knows when to remove 
 the seed from its memory.
 Pheromone seeds are therefore forwarded as messages of the form:
 $(P,c,f,a,t)$ (pheromone seed, counter, remaining fees,  payment amount and timestamp).\\
 
   For each matched seed $M$, the node stores the following items:
   
   \begin{enumerate}[(1)]
  \item A ``match identifier'' $Id$.
  \item The ``target'' of the match, which is the name of the next node in the path from Alice to Bob which $M$ indicates.
 \end{enumerate}

 The number $Id$ is a random number generated at the creation of the match. Its purpose is to distinguish between different matches 
 which may have been created from the same pheromone seed. Note that it is not necessary to store $M$ itself, although $M$ needs to be 
 forwarded. There are also four extra pieces of information which need to be forwarded but do not have to be kept in memory. These are: 
 
 \begin{enumerate}[(1)]
  \item Two counters $c$ and $C$. The counter $C$ gives the number of intermediate nodes from Alice to Bob in the path indicated by $M$.  
  The counter $c$ is decreased by $1$ at each hop. Its purpose is to be compared with the counter of 
  the pheromone seed associated to $M$ (see section \ref{algorithmsection} below).
  \item The total remaining fees $F$ associated to $M$. We have 
  $F=2f_{max}-G$, where $G$ denotes
  the sum of all fees from Alice to Bob 
  in the path indicated by $M$. The purpose of $F$ is to be transmitted to Alice, so that she can choose the path with the 
  lowest fees.
  \item A timestamp $t$, which is the timestamp of the corresponding pheromone seed.
 \end{enumerate}

 Therefore, matched seeds are forwarded to other nodes via a message of the form 
 $(M,Id,c,C,F,t)$. Note that, in Alice's special match tree, $F$ and $C$ are also stored.\\
 
 A confirmation seed is simply a random number $Id$. For each confirmation seed, 
 the node stores:
    \begin{enumerate}[(1)]
     \item An identifier $Id$. This is the identifier of the matched seed chosen by Alice.
     \item The ``target'', i.e. the next node in the path from Alice to Bob.
     \item An integer $check$. This is a randomly generated integer meant for the final counter check phase.
    \end{enumerate}
  Confirmed seeds are forwarded in messages of the form 
 $(Id,l,t)$, where $l$ is a list of integers and $t$ is the timestamp of the corresponding pheromone seed. More precisely, $l$ is the list of the ``$check$'' integers of all preceding nodes.\\

 \begin{Rem}
 The node does not store the timestamp of a given 
 seed. However,
 the location of a given seed  in memory depends on its timestamp $t$. Indeed, we will see in Section \ref{Implementationsection} that 
  seeds are stored in several trees, each of which corresponds to a time interval to which the seed's timestamp belongs. 
 It follows that, in order to find a seed in its memory, the node needs to know the associated timestamp. 
 This explains why the timestamp also has to be forwarded with each type of seed.
\end{Rem}
 
 \subsection{Algorithms}\label{algorithmsection}
   The ant routing protocol has four phases: the ``pheromone'' phase, the ``match'' phase, the 
   ``confirmation'' phase and the ``counter check'' phase. We now describe these phases one by one in this order.

 Alice chooses  two random numbers $S$ and $c_0$, the timestamp $t$, and the maximal amount of fees 
 $f_{max}$ which she is willing to pay, and communicates to Bob this data.
 The number 
 $S$ will be used to form the pheromone seeds and  $c_0$ will be used the starting value for the counter\footnote{The value of $c_0$ can be chosen 
 between $2^6=64$ and $2^7=128$. By doing this, we only dedicate 1 Byte to the counter, but still allow for $128$ intermediary nodes between Alice 
 and Bob, which is enough.}.
 The purpose of choosing a random number as the starting value of the counter  is to preserve Alice's and Bob's anonymity.
 If we start the counter at $0$, then the neighbors of ALice and Bob will know that they are originating the transaction. 
 The timestamp $t$ serves as the time at which Alice and Bob broadcast their pheromone seeds.

   Assume that $S,c_0$ and $t$ have been agreed upon by Alice and Bob, and that Alice has chosen $f_{max}$.
    Alice then creates $P(0):=0^{\frown}S$ and stores the information 
    $(P(0),c_0,0,f_{max},a)$ in her pheromone tree, where $a$ denotes the amount of money which she wants to send Bob. The  ``$0$'' indicates that the seed 
    has no sender, so the seed originates from Alice. 
    Similarly, Bob creates $P(1):=1^{\frown}S$ and stores the information
    $(P(1),c_0,0,f_{max},a)$ in his pheromone tree.    
    At time $t$,
    Alice sends the message $(P(0),c_0,f_{max},a,t)$ to all her neighbors,
    and 
    Bob sends the message $(P(1),c_0,f_{max},a,t)$ to all his neighbors.\\
    
    After this initial step, the nodes will keep forwarding $P(0)$ and $P(1)$ until matches occur, increasing the counter by $1$ 
   at each hop.
 In general, the same node will receive the same pheromone seed several times coming from different paths. However, after a pheromone seed has 
 been received the first time, then the node will only broadcast new arrivals of the same seed if the counter of the newly arrived seed is 
 smaller than the counter of the previously received seed. This reduces the amount of information sent through the network and makes sure that 
 Alice only receives the shortest route proposals. 
 
    Each node substracts its own fees from the remaining fees before forwarding the seed. A node only has knowledge of remaining fees, 
    and does not know $f_{max}$. This makes it extremely difficult for intermediary nodes to deduce 
    from the fees
    any information about the sender or receiver of the payment, thus preserving Alice's and Bob's anonymity.
 
 Here is an algorithm describing the work of a node upon receiving a pheromone seed from node $s$
 in the form of a message $(P,c,f,a,t)$. We denote by $g$ the fees of the node performing the task. $\overline{P}$ denotes the conjugate of $P$, i.e $\overline{P(0)}=P(1)$ and $\overline{P(1)}=P(0)$.\\

    \begin{algorithm}[H]\label{pheromonealgo}
    \caption{treatment of pheromone seeds}
   Look for $P$ in memory;
   
  \uIf{$P$ is not in memory and $f-g\geq0$}{  
  
    Insert $(P,c,s,f,a)$ in memory;    
  }     
     \uElseIf{$(P,c',s',f',a)$ is in memory}{
      \uIf{$c'\leq c$}{
      Exit program;
      }

      \uElseIf{$f-g\geq0$}{ 
      Replace $(P,c',s',f',a)$ by $(P,c,s,f,a)$ in memory. 
      }}

  Look for  $\overline{P}$ in memory.
  
  \uIf{$\overline{P}$ is not in memory}{
      Send ($P,c+1,f-g,a,t$) to all      
     neighbors with channel balance at least $a$, except $s$. 
     }

     \uElse{
     $create\_and\_send\_match(P)$.
     }
       
 \end{algorithm}

 \vspace{10pt}

 The function $create\_and\_send\_match()$ used above starts the match phase of the ant routing protocol.
 A matched seed has the form $M(0):=0^{\frown}P(0)$ or $M(1):=0^{\frown}P(1)$, where $P(0)$ and $P(1)$ are pheromone seeds. 
 $M(0)$ is sent back to Alice, and $M(1)$ is sent back to Bob.
 Assume $(P(0),c,s,f,a)$ and $(P(1),c',s',f',a)$ are both in memory, and denote by $g$ the fees of the node where the match occurs. Then 
 the function
 $create\_and\_send\_match(P)$ does the following:\\
 
 \begin{algorithm}[H]\label{creatematchalg}
 \caption{Match creation}
 \uIf{$f+f'-g\geq0$}{
  Set $M(0):=0^{\frown}P(0)$, $M(1):=0^{\frown}P(1)$.
  
  Generate a random number $Id$;
  
  Set $F:=f+f'-g$;
  
  Set $C:=c+c'+1$;
  
  Store $(Id,s')$ in the match tree;
  
  Send  $(M(0),Id,c,C,F,t)$ to $s$;
  
Send $(M(1),Id,c',C,F,t)$ to $s'$;
  }
 \end{algorithm}
 \vspace{10pt}

 Matched seeds are forwarded as messages of the form $(M(\epsilon),Id,c,C,F,t)$, $\epsilon\in\{0,1\}$. The counter $c$ is decreased by $1$ at each hop. 
 When a node receives a matched seed $(M(\epsilon),Id,c,C,F,t)$, it looks for the corresponding 
 $P(\epsilon)$ in its memory and checks that the counter associated to $P(\epsilon)$ is equal to $c-1$. 
 If this is the case, then the node retrieves the sender $s$ of $P(\epsilon)$ and forwards $M(\epsilon)$ to 
 $s$. 
 If the counter associated to $P(\epsilon)$ does not agree with $c-1$, this means 
 that, before receiving $M$, the node has received a new instance of $P(\epsilon)$ with lower counter. But then the information associated to $P(\epsilon)$ in the node's memory do not correspond 
 to the information associated to $M(\epsilon)$, and thus $M(\epsilon)$ must be discarded. However, the node will eventually
 receive a new instance of 
 $M(\epsilon)$ with new information.

 Here is the algorithm for the treatment of matched seeds. The input is a message of the form 
 $(M(\epsilon),Id,c,C,F,t)$ received from node $y$, and $g$ denotes the fees of the node performing the task:\\
 
 \begin{algorithm}[H]
 \caption{Treatment of matched seeds}
   \uIf{$\epsilon=0$}{
 Look for $P(0)$ in memory.
 
 \uIf{$(P(0),c',s,f,a)$ is in memory}{   
  
  \uIf{$c'\neq c-1$}{
    Exit program;}
  \uIf{$s=0$}{
  Store $(Id,y,C,F)$ in the  special match tree}
  \uElse{
   Store $(Id,y)$ in the general match tree;
   
   Send $(M(\epsilon),Id,c-1,C,F)$ to $s$. }
  }}

  \uIf{$\epsilon=1$}{
  
  Look for $P(1)$ in memory.
  
  \uIf{$(P(1),c',s,f,a)$ is in memory}{   
 \uIf{$c'\neq c-1$}{
    Exit program;} 
   
   Store $(Id,s)$ in memory.
   
 \uIf{$s\neq0$}{
   
   Send $(M(\epsilon),Id,c-1,C,F,t)$ to $s$.}
  }}
 \end{algorithm}
 \vspace{10pt}
 
 After some time, Alice will have received several matched seeds, stored in her special match tree.
 She now chooses one (say, the one with the lowest fees), which is stored as $(Id,y,C,F)$. Note that Alice can 
 recover the amount of fees she will have to pay by computing $2f_{max}-F$. This number is bounded by $2f_{max}$, but not by 
 $f_{max}$. If Alice is not willing to pay up to $2f_{max}$ in fees, then she can choose to replace $f_{max}$ by 
 $\frac{f_{max}}{2}$ at the beginning of the pheromone phase.
 
 To launch the confirmation phase,
 Alice generates a list of random numbers $l_0$, which will serve for the final counter check.
  Alice then sends the confirmation seed $(Id,l:=l_0,t)$ to $y$. This starts the confirmation phase of the ant routing protocol. 
 
 Here is the algorithm performed by a node after receiving the confirmation seed $(Id,l,t)$:\\
 
 \begin{algorithm}[H]\label{confirmationalgo}
 \caption{Treatment of confirmed seeds}
 
  Look for the match identifier $Id$ in the match tree;
  
  \uIf{$(Id,s)$ is in the match tree and $s\neq0$}{

    Generate a random integer $check$.
    
    Append $check$ to $l$.
    
    Store $(Id,s,check)$ in the confirmation tree;
    
    Send $(Id,l,t)$ to $s$;
    }

 \end{algorithm}

Remember that the ``$s$'' in the matched seed $(Id,s,t)$ refers to the sender of the pheromone seed from which $Id$ was made, and 
that $s=0$ only for Bob and Alice. Therefore, the ``$s\neq0$'' clause in algorithm \ref{confirmationalgo} 
states that the algorithm is executed by every intermediary node but not by Bob.
When Bob receives $(Id,l,t)$, then Bob sends $l$ to Alice. 
Alice then checks that the number of random numbers which were appended to $l$ is equal to $C-2c_0$ (note that the number of intermediary nodes 
between Alice and Bob is $C-2c_0$). 
If this is not the case, this means that one of the nodes on the route indicated by $Id$ is a cheater. Alice then chooses another route: she chooses  another ``$Id$'' seed in her special match memory and starts a new confirmation phase with this new ``$Id$''.

If the number of random numbers appended to $l$ matches $C$, then Alice removes $l_0$ from 
$l$.  She appends a new list of random numbers $l_1$ at the end of $l$ and sends $(Id,l,t)$ to $y$. Then each node on the path checks that the first number in $l$ is the one that they generated,
i.e each node follows the following algorithm:\\

\begin{algorithm}[H]
\caption{counter check round}
 Look for $Id$ in the confirmation tree;
 
 \uIf{$(Id,s,check)$ is in the confirmation tree and $s\neq0$}{

    \uIf{$check=l[0]$}{
    Remove $check$ from $l$;
    
    Send $(Id,l,t)$ to $s$;
    }
    }
 
\end{algorithm}

 When Bob receives $(Id,l,t)$, he sends a message to Alice telling her that she can proceed with the payment.  Note that the list $l_1$ that has been appened to $l$ at the start of the counter check round is here so that nodes don't know the lenght of the payment route. In particular, no node knows that Alice is paying Bob.

 \begin{Rem}
  One way of reducing the workload on the network would be to use a variant of the ant routing algorithm which we just presented. 
   In this variant, only Alice, and not Bob, sends pheromone seeds. Bob then waits until he receives the pheromone seed from Alice.
   Bob may receive several seeds, each indicating a different route. Bob can then communicate the information he has on each of these routes  to Alice, who then chooses one of them. Bob then creates the match himself from the pheromone seed chosen by Alice, 
   and sends the matched seed back to Alice though the chosen route. 
   
   Note that this variant is only viable if Alice trusts Bob. Indeed, since Bob knows the remaining fees for each seed he receives, he could cheat Alice by taking all the remaining fees for himself.
 \end{Rem}

 \section{Robustness of ant routing}
 
 Payment routing depending on knowledge of the network threatens decentralization. 
 If knowledge of the network is required for payment, some nodes may choose to delegate the task of gathering knowledge to 
 other nodes. This can lead to the emergence of ``hub'' nodes which centralize most routing tasks, with most nodes at the periphery. 
 In particular, the use of landmarks encourages this process. This type of network is easy to disrupt: if all payments depend on a small
 group of nodes, then one only needs to neutralize these nodes to incapacitate the whole network.
 
 An important advantage of the ant routing algorithm is its total decentralization. All nodes in the network performs the exact same 
 task and obey the exact same rules. To have the same protocol rules for all nodes is an important key idea of decentralization. 
 Ant routing requires no knowledge of the geometry of the network, which prevents the emergence of ``hub'' nodes.
 This makes the network particularly resilient to attacks: even if an attacker neutralizes a larger part of the network, 
 the remaining part can still function normally provided it remains connected.
 
 It is worth noting that the main reason for the high failure rate of the current routing algorithm in the lightning network is the fact that 
 nodes who compute payment routes only know the capacities of the channels on the network, but they do not know their balances. As a consequence, the paying node often ends up choosing a route which cannot relay the payment due to insufficient available funds in one direction \cite{NP19}. 
 Ant routing solves this problem by making sure that every channel on the path has sufficient funds in the desired direction. Indeed, during the pheromone phase, a node will forward the seed to a neighbor only if the funds available on their channel is sufficient. As a consequence, Alice 
 knows that, for every match which she receives, the channels on the path indicated by the match have enough funds to forward her payment.
 
   Note also that ant routing addresses some concerns about the privacy of lightning transactions which have been raised in 
   \cite{Beres}. The authors of \cite{Beres} noticed that many LN transactions only have one intermediary node between payer and payee, which means that the 
   intermediary node knows their identities. Ant routing solves this problem by offering several possible path to Alice. Thanks to the counter, Alice knows the length of each path indicated by a match, and she can choose a path of length at least 3 if she wants to preserve her privacy.

A malicious node may attempt to disturb the network by cheating on the counter. Indeed, a malicious node may add a 
negative counter to a pheromone seed to make the path  more attractive. Setting very small fees also increases the chances 
of Alice choosing this path. The malicious node could then refuse to transmit the payment, thus disturbing the network. 
 Note however that, thanks to the counter check phase which we added at the end of ant routing, Alice will discover the presence 
 of a malicious node before sending the payment. Because the ant routing algorithm usually returns several matched seeds to Alice, she can then choose another 
 route, avoiding the malicious node. If Alice sees that the confirmation phase, the counter check phase or the payment itself did not 
 terminate successfully, she can send a message to the other nodes of the path to warn them of the presence of a malicious node. 
 If a node notices frequent problems with attempted payments through one of its neighbors, it can decide to stop forwarding pheromone 
 seeds throught him or even to close the payment channel. In this way, the
 malicious node will eventually be de facto isolated and rejected from the network, and will not be able to disturb it any longer. 
 The same will happen with a denial of service attack by nodes that respect all the rules 
 of the protocol but in the last moment refuse to collaborate 
 in establishing the payment. Therefore, each node has interest in keeping historical data and 
 statistics on the performance of its neighbors to avoid 
 being connected to dishonest nodes. In this sense, the network behaves more like an ant 
 colony since the traces left by older payments will reinforce the best neighbors.

\section{Life time of seeds}\label{lifetimesection}

The seeds are only useful for the routing task for a transaction, and must therefore be kept only a short time in memory. 
This raises the question of the life time of seeds: how long should the seeds be kept in memory? 
They need to be kept long enough to allow the routing task to be completed, but as short as possible 
in order to reduce memory usage.

 To determine the optimal seed life time, we need an estimate of how long it will take to  
 Alice to receive matched seeds. This depends on the bandwith for data propagation on the Lightning Network. 
 We can have an approximate estimate using the Bitcoin network as a proxy for the lightning network. Both networks are 
 topologically of the same nature, more precisely, they are well-connected networks. Also seeds in ant 
 routing do propagate faster through the network than bitcoin transaction on the Bitcoin network. This is 
 due to the fact that pheromone seeds are much smaller than bitcoin transactions, but also because the local node 
 process of transactions in Bitcoin nodes is much heavier than the process of pheromone seeds. It follows that the speed 
 of transaction propagation on the Bitcoin network gives us a generous upper 
 bound on the speed of pheromone propagation on the Lightning Network.
 
   We found on \cite{datapropagation} that, in 2014, it took around $0.8$ seconds for a transaction 
   to reach $50\%$ of the nodes in the Bitcoin network. This is in accordance with \cite{Erlay} (See their fig. 16). It follows 
   that we make the conservative assumption that in average a pheromone seed reaches $50\%$ of nodes in $0.8$ seconds. Assuming the 
   initial independence of propagation of Alice and Bob seed, for a given node, the probability to have a match after
   $0.8$ seconds is $0.5^2=0.25$ and the probability that there is a match somewhere in the network after $0.8$ seconds is
   $1-0.75^N$ where $N$ is the number of nodes. For $N\geq 32$,
   this probability is more than $99.99\%$.
   Now, if we assume that both Alice's pheromone and Bob's pheromone have reached $10\%$ of the nodes in $0.5$ seconds, then, the probability to have a match at this date is $1-0.99^N$ which is greater than $99.99\%$ when $N\geq 1000$ which is already the case today.
The exact distribution of the time it takes to have a first match depends on the network topology, but in view 
of these numbers it is reasonable to assume the expected unmatched time cycle of a pheromone seed to be about $0.5$ seconds. Obviously, this may be higher when Alice or Bob do have a bad connectivity to the network, but this situation is neglectable if the average node is well connected as we assume. Note also that this estimate implies that the expected duration time for the whole routing task to be completed is less than $1$ second.

\section{A general maximum capacity estimate}\label{proxymodelsection}

We can develop the previous argument using the Bitcoin network as a proxy for estimating the  maximum network capacity of an arbitrary network with similar 
characteristics. We first estimate in an elementary way the maximum network capacity of the Bitcoin network. 
We assume instant propagation of the data across the network and that the local nodes workload time for the 
propagation of a transaction is neglectable. The estimate remains valid if there is no lag caused by the local workload of nodes. 
We consider several random variables, for $t\geq 0$,
\begin{itemize}
 \item $\boldsymbol{N}(t)$ number of blocks mined by the network at time $t$, $\boldsymbol{N}(0)=0$.
 \item $\boldsymbol{L}(t)$ number of transactions created by the network at time $t$, $\boldsymbol{L}(0)=0$.
 \item $\boldsymbol{M}(t)$ mempool size in Bytes at time $t$, $\boldsymbol{M}(0)=M_0$.
 \item $\boldsymbol{B} (b)$  size in Bytes of the  block number $b$.
 \item $\boldsymbol{m}(x)$ size in Bytes of transaction $x$. 
 \item $\boldsymbol{t}(x)$ time when transaction $x$ was created. 
\end{itemize}
Note that $\boldsymbol{L}, \boldsymbol{M}, \boldsymbol{N}$ are random process, and $\boldsymbol{L}, \boldsymbol{N}$ are Poisson processes \cite{GrunspanPerez2017}.
The distribution of $\boldsymbol{M}(t)$ depends on the network condition. 
At $t=0$ we start counting blocks $B_1, B_2, \ldots$ and transactions created $x_1, x_2,\ldots$ from this 
moment and the initial condition $M(0)=M_0$. We have
$$
M(t)= M_0+\sum_{k=1}^{\boldsymbol{L}(t)} m(x_k) -\sum_{k=1}^{\boldsymbol{N}(t)} B(b_k) \ .
$$
In the steady regime of maximum capacity, $\boldsymbol{M}(t)$ is constant, $\boldsymbol{M}(t)=M_0$. Taking expected values in the precedent 
equation, we get, using Wald's Theorem,
$$
0=\mathbb{E}[\boldsymbol{L}(t)] \mathbb{E}[\boldsymbol{m}]- \mathbb{E}[\boldsymbol{N}(t)]  \mathbb{E}[\boldsymbol{B}]
$$
thus
$$
\mathbb{E}[\boldsymbol{L}(t)] =\frac{\mathbb{E}[\boldsymbol{N}(t)]  \mathbb{E}[\boldsymbol{B}]}{\mathbb{E}[\boldsymbol{m}]}
$$
We have, $\mathbb{E}[\boldsymbol{N}(t)]=t/\tau_0$ where $\tau_0$ is the average inter-block time.
At steady maximum capacity regime $\mathbb{E}[\boldsymbol{B}]$ is constant and equal to the maximum block size, $\mathbb{E}[\boldsymbol{B}]=B_{\text{max}}$. Let 
$m_0=\mathbb{E}[\boldsymbol{m}]$ the expected size of a transaction. These calculations show the following result:
\begin{Thm}
With the previous assumptioons, at steady maximum capacity regime for typical transactions we have
$$
l_0 =\frac{B_{\text{max}}}{m_0 \tau_0}
$$
where $l_0$ is the maximum number of transactions per second, $B_{\text{max}}$ is the maximum block size, 
$m_0$ is the average size of a transaction, and $\tau_0$ is the average inter-block time.
\end{Thm}

The arguments given to derive the above formula are quite general and apply to any network with similar characteristics as the Bitcoin network.

\begin{Application*}[Bitcoin network]
We can apply the precedent formula to the Bitcoin network. We have $\tau_0=10\text{ min}$ (slightly smaller when difficulty is increasing). Also, 
before segwit improvement and neglecting the block header size of $80 B$, 
$B_{\text{max}}=1\text{ MB}$. Typical Bitcoin transactions have a size of $m_0=250 \text{ B}$. We remind that one bitcoin transaction can contain  
several financial transactions and the size depends sensitively in the number of inputs and outputs. From this we get 
$$
l_{\text{BTC}} = 6.67 \text{ tx/sec}
$$ 
for the maximum number of typical transactions per second in the Bitcoin network. The minimal size of a transaction is $61 \text{ B}$ (or $63 \text{ B}$ for a coinbase transaction, but this is neglectable, see \cite{Gio}) then we get
$$
l_{\text{BTC}} = 27.3 \text{ tx/sec}
$$ 

\end{Application*}

\begin{Application*}[Monero network]
Our formula applies directly. The maximum block size is set to $M_0=1 \text{ MB}$, the current average size of a transaction is $m_0=550 \text{ B}$ and 
the average interblock time of $\tau_0=120 \text{ sec}$. Our formula gives
$$
l_{\text{XMR}} = 15.15 \text{ tx/sec}
$$
for the maximum number of transactions per second in the Monero network. 
\end{Application*}

\begin{Application*}[Ethereum network]
We apply the precedent ideas to the Ethereum network. The Ethereum protocol imposes no size limit on blocks or transactions, but the miners set a gas limit 
$G_{\text{max}}$ for the blocks, currently set at $G_{\text{max}}=10^7$ units of gas, that is the total sum of gas for the transactions in the block. 
The gas of a transaction roughly measures the complexity of the computations in the smart contracts 
in Ethereum transactions. Indeed one line of code costs $1$ gas unit, hence there is some correlation between the gas of a transaction and its size. 
But the proper metric to measure the network capacity is by using our formula where we replace the $B_{\text{max}}$ by $G_{\text{max}}$, and $m_0$ by the 
average gas $g_0$ required by a transaction. The formula for the maximum number of transactions per second is 
$$
l_{ETH} =\frac{G_{\text{max}}}{g_0 \tau_0}
$$
We currently have $\tau_0=15 \text{ sec}$.
Observe that the full capacity of the network can be reached by a maximum of basic transactions with minimal gas, or by a few transactions with maximal gas.
For $g_0 =g_{\text{min}} =21.000$ we have
$$
l_{ETH}=31.75 \text{ tx/sec}
$$
For $g_0 = g_{\text{max}} =200.000$ we have
$$
l_{ETH} =3.33 \text{ tx/sec}
$$
Currently, the observed historical maximum block size, according to the block explorer Etherscan, is $36\, 630 \text{ B}$. Note that this represents $1.46 \text{ MB}$ size per $10 \text{ min}$.
The average Ethereum transactions size is about  $m_0=500 \text{ B}$ (not counting data volume necessary for the smart contract). with these figures, applying 
our formula we get 
$$
l_{ETH} = 4.88 \text{ tx/sec}
$$
for the maximum number of transaction per seconds assuming a maximum size block of $36\, 630 B$. Obviously this size limit is not set by the protocol, but the 
practical limit has the same order of magnitude. Ethereum has similar scaling problems than Bitcoin, and this is reflected by  the 
current heavy size of its full blockchain.
\end{Application*}

\begin{Application*}[Ant Routing]
The same argument leading to the formula applies to the Lightning Network running the Ant Routing protocol. This time we denote by $M_0$ is the maximum size of the 
local mempool at each node. Let  $\mu_0$ the average size of data propagated through all the network for each transactions, essentially unmatched and 
matched pheromone seeds so $\mu_0 \simeq 100 \text{ B}$. Let $\eta$ the average life time of most data in the Lightning network, we can take the conservative 
figure of $\eta=2 \text{ sec}$. We have,
$$
l_{AR} = \frac{M_0}{\tau_0\eta}
$$
for the number of transactions per second of the Lightning Network running on Ant Routing. For the conservative figures choosen, we have
$$
l_{AR} = \frac{M_0}{200}
$$
For example,  for a modest mempool of $M_0=20 \text{ MB}$ we get 
$$
l_{AR} = 10.000 \text{ tx/sec}
$$
As we will see in the next sections,  the numerical simulations of the local process time 
of seeds at nodes allow for a flow of $10^6 \text{ tx/sec}$ with no  impact on the propagation and matching time, thus the 
network can process an order of $10^6 \text{ tx/sec}$.

\end{Application*}

 \section{AVL trees}\label{AVLsection}
 In this section we describe the structure we choose in later sections 
 to implement the ant routing protocol: AVL trees. AVL trees were introduced in \cite{AdelsonLandis}
 
 We first fix some notations. 
 If $T$ is a binary tree, we denote by 
 $h(T)$ the height of $T$. If $x$ is a node of $T$, we denote by 
 $T_x$ the subtree of $T$ generated by $x$. $T_x^l$ denotes the subtree generated 
 by the left child of $x$, and $T_x^r$ denotes the subtree of $T$ generated by the right child of $x$. If $x$ has no left child (respectively, right child), we say that its left child (respectively, right child) is NULL.\\
 
  A \textbf{binary search tree} is a binary tree $T$ such that: 
 \begin{enumerate}
  \item Every node of $T$ is labelled with an integer.
  \item For every nodes $x,y$ of $T$ respectively labeled by the integers 
  $n$ and $m$, we have 
  $y\in T_x^l\Rightarrow m<n$ and 
  $y\in T_x^r\Rightarrow m>n$.
 \end{enumerate}

  If $T$ is a binary search tree, we say that $T$ is  \textbf{balanced}
  if for every $x\in T$,\newline  
  $\vert h(T_x^l)-h(T_x^r)\vert\leq1$.
  An \textbf{AVL tree} is a balanced binary search tree.
  
  The following is a look-up algorithm for a binary search tree. It takes 
  an integer $n$ as entry, and returns the node of $T$ labeled by $n$ if it exists, and returns NULL if no node of $T$ is labeled by $n$.\\
  
  \begin{algorithm}[H]\label{lookupalgo}
\SetAlgoLined
\SetKwInOut{Input}{input}\SetKwInOut{Output}{output}

 \Input{Integer $n$, tree $T$}
\Output{$lookup(n,T)$:=the node of $T$ labeled by $n$ if it exists, ``NULL'' otherwise}
 Set $node:=root(T)$;
 
  \While{$node\neq NULL$}{
  
  \uIf{$n<label(node)$}{
  $node:=left\_child(node)$;
 }
   
   \uElseIf{$n>label(node)$}{
   $node:=right\_child(node)$;
  }
  \uElse{
 return $node$;}
 }
 Return $node$;
 \caption{Look-up algorithm for binary search tree.}
\end{algorithm}
\vspace{10pt}

Once the lookup is done, insertion in a binary search tree can be done fast. Here is an algorithm which 
inserts $n$ in $T$ if $n$ is not already in $T$:

\begin{algorithm}[H]\label{Insertionalgo}
\SetAlgoLined
\SetKwInOut{Input}{input}\SetKwInOut{Output}{output}

 \Input{Integer $n$, tree $T$}
 Set $node=root(T)$;
 
  \While{$true$}{
  
  \uIf{$n<label(node)$}{
  \uIf{$left\_child(node)=NULL$}{
    $left\_child(node):=create\_node()$;
    
    $label(left\_child(node)):=n$;
    
    Exit program;}
 \uElse{
    $node:=left\_child(node)$;}
 }
   
   \uElseIf{$n>label(node)$}{
  \uIf{$right\_child(node)=NULL$}{
    $right\_child(node):=create\_node()$;
    
    $label(right\_child(node)):=n$;
    
    Exit program;}
 \uElse{
    $node:=right\_child(node)$;}
 }
  \uElse{
 Exit program;}
 }
 \caption{Insertion algorithm for binary search tree.}
\end{algorithm}

  Now we discuss  the run time of these algorithms. Let $N$ be the number of nodes in $T$. 
  If $T$ is not assumed to be balanced, then in the worst case algorithms
  \ref{lookupalgo} and \ref{Insertionalgo} could run in time $O(N)$. However, if 
  $T$ is an AVL tree, then algorithms \ref{lookupalgo} and \ref{Insertionalgo} are guaranteed to run in 
  time $O(\log_2(N))$.

  Algorithm \ref{Insertionalgo} is a naive insertion algorithm. This naive algorithm could be problematic, as it takes no account of the height of the newly generated tree and could thus result in an unbalanced tree if repeated enough times.
  This means we cannot use algorithm \ref{Insertionalgo} for insertion in AVL trees. Fortunately, there is a workaround to this problem. More precisely, we can slightly change the insertion algorithm so that the tree remains balanced. 
  \setcounter{figure}{1}
  We now describe the insertion algorithm for AVL trees.
  We first need to explain how an unbalanced tree can be rebalanced.
  
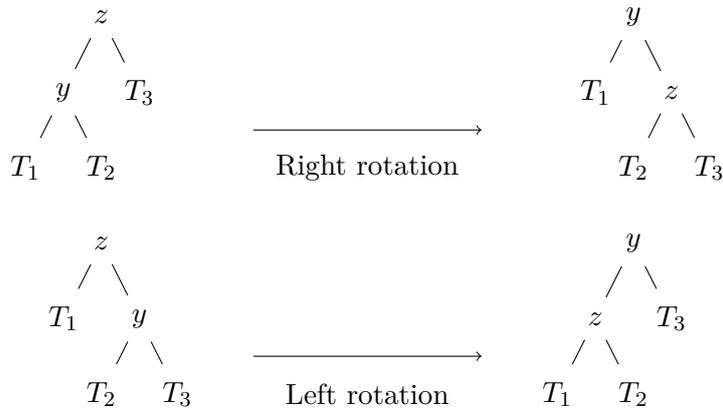
\begin{figure}[H]
  \caption{Rotations at node $z$.}\label{rotations}
  \begin{tikzpicture}
  \tikzstyle{noeud}=[circle]
  
   \node[noeud] (z) at (0,0) {$z$};
   \node[noeud] (T3) at (0.5,-1) {$T_3$};
   \node[noeud] (y) at (-0.5,-1) {$y$};
    \node[noeud] (T2) at (0,-2) {$T_2$};

   \node[noeud] (T1) at (-1,-2) {$T_1$};

   \draw (z)--(y);
   \draw (z)--(T3);
   \draw (T1)--(y);
   \draw (T2)--(y);

   \node[noeud] (y') at (7,0) {$y$};
   \node[noeud] (z') at (7.5,-1) {$z$};
   \node[noeud] (T3') at (8,-2) {$T_3$};
   
    \node[noeud] (T2') at (7,-2) {$T_2$};

   \node[noeud] (T1') at (6.5,-1) {$T_1$};

   \draw (z')--(y');
   \draw (z')--(T3');
   \draw (T1')--(y');
   \draw (T2')--(z');

   \draw[->] (2,-1.5)--(5,-1.5); 
   \node[noeud] (text) at (3.5,-2) {Right rotation};

   \node[noeud] (zz) at (0,-3) {$z$};
   \node[noeud] (yy) at (0.5,-4) {$y$};
   \node[noeud] (TT1) at (-0.5,-4) {$T_1$};
    \node[noeud] (TT2) at (0,-5) {$T_2$};

   \node[noeud] (TT3) at (1,-5) {$T_3$};

   \draw (zz)--(yy);
   \draw (yy)--(TT3);
   \draw (TT1)--(zz);
   \draw (TT2)--(yy); 
   
   \node[noeud] (y'') at (7,-3) {$y$};
   \node[noeud] (z'') at (6.5,-4) {$z$};
   \node[noeud] (T3'') at (7.5,-4) {$T_3$};
   
    \node[noeud] (T2'') at (7,-5) {$T_2$};

   \node[noeud] (T1'') at (6,-5) {$T_1$};

   \draw (z'')--(y'');
   \draw (y'')--(T3'');
   \draw (T1'')--(z'');
   \draw (T2'')--(z'');

   \draw[->] (2,-4.5)--(5,-4.5); 
   \node[noeud] (text) at (3.5,-5) {Left rotation};
  \end{tikzpicture}
\end{figure}

  Re-balancing a tree is done via operations on the tree called rotations. A rotation 
  at a node $z$ is one of the two operations shown in Figure \ref{rotations}.

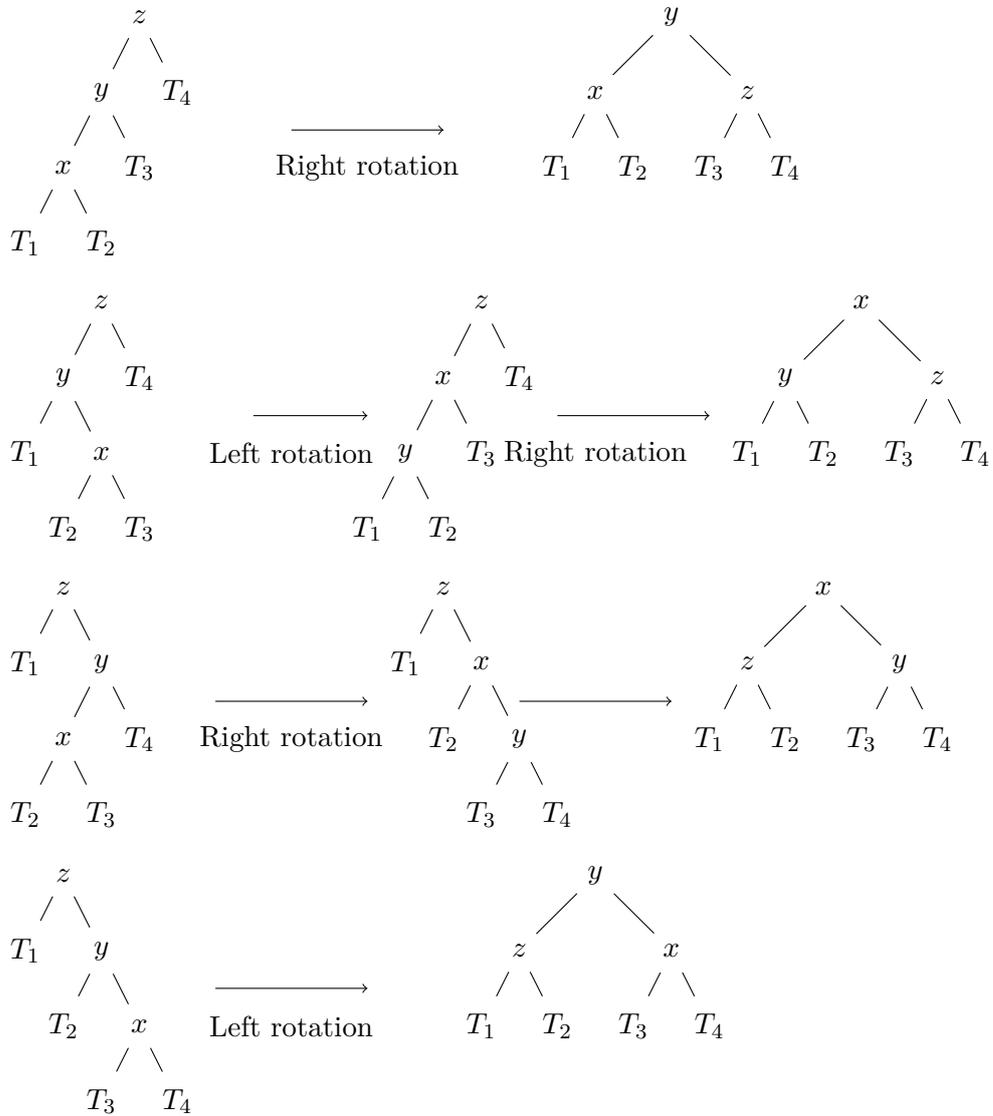
\begin{figure}[H]
  \caption{Rebalancing the tree}\label{rebalancefigure}
  
  \begin{tikzpicture}
  
  \tikzstyle{noeud}=[circle]
  
   \node[noeud] (z) at (0,0) {$z$};
   \node[noeud] (T4) at (0.5,-1) {$T_4$};
   \node[noeud] (y) at (-0.5,-1) {$y$};
    \node[noeud] (T3) at (0,-2) {$T_3$};

   \node[noeud] (x) at (-1,-2) {$x$};
   \node[noeud] (T1) at (-1.5,-3) {$T_1$};
    \node[noeud] (T2) at (-0.5,-3) {$T_2$};

   \draw (z)--(y);
   \draw (z)--(T4);
   \draw (x)--(y);
   \draw (T3)--(y);
   \draw (x)--(T1);
   \draw (x)--(T2);

   \node[noeud] (y') at (7,0) {$y$};
   \node[noeud] (z') at (8,-1) {$z$};
   \node[noeud] (T4') at (8.5,-2) {$T_4$};
   
    \node[noeud] (T3') at (7.5,-2) {$T_3$};

   \node[noeud] (x') at (6,-1) {$x$};
   \node[noeud] (T1') at (5.5,-2) {$T_1$};
    \node[noeud] (T2') at (6.5,-2) {$T_2$};

   \draw (z')--(y');
   \draw (z')--(T4');
   \draw (x')--(y');
   \draw (T3')--(z');
   \draw (x')--(T1');
   \draw (x')--(T2');
   
   \draw[->] (2,-1.5)--(4,-1.5); 
   \node[noeud] (text) at (3,-2) {Right rotation};
   
  \end{tikzpicture}

\begin{tikzpicture}
\tikzstyle{noeud}=[circle]
 \node[noeud] (z) at (0,0) {$z$};
   \node[noeud] (T4) at (0.5,-1) {$T_4$};
   \node[noeud] (y) at (-0.5,-1) {$y$};
    \node[noeud] (T3) at (0.5,-3) {$T_3$};

   \node[noeud] (x) at (0,-2) {$x$};
   \node[noeud] (T1) at (-1,-2) {$T_1$};
    \node[noeud] (T2) at (-0.5,-3) {$T_2$};

   \draw (z)--(y);
   \draw (z)--(T4);
   \draw (x)--(y);
   \draw (T3)--(x);
   \draw (y)--(T1);
   \draw (x)--(T2);
   
   \node[noeud] (z') at (5,0) {$z$};
   \node[noeud] (T4') at (5.5,-1) {$T_4$};
   \node[noeud] (x') at (4.5,-1) {$x$};
   \node[noeud] (T3') at (5,-2) {$T_3$};
   \node[noeud] (y') at (4,-2) {$y$};
    \node[noeud] (T2') at (4.5,-3) {$T_2$};
    \node[noeud] (T1') at (3.5,-3) {$T_1$};

   \draw (z')--(x');
   \draw (z')--(T4');
   \draw (x')--(y');
   \draw (T3')--(x');
   \draw (y')--(T1');
   \draw (y')--(T2');
   
   \draw[->] (2,-1.5)--(3.5,-1.5); 
   \node[noeud] (text) at (2.5,-2) {Left rotation};

   \node[noeud] (x'') at (10,0) {$x$};
   \node[noeud] (z'') at (11,-1) {$z$};
   \node[noeud] (y'') at (9,-1) {$y$};
   \node[noeud] (T3'') at (10.5,-2) {$T_3$};
   \node[noeud] (T4'') at (11.5,-2) {$T_4$};
    \node[noeud] (T2'') at (9.5,-2) {$T_2$};
    \node[noeud] (T1'') at (8.5,-2) {$T_1$};

   \draw (z'')--(x'');
   \draw (z'')--(T4'');
   \draw (x'')--(y'');
   \draw (T3'')--(z'');
   \draw (y'')--(T1'');
   \draw (y'')--(T2'');
   \draw[->] (6,-1.5)--(8,-1.5); 
   \node[noeud] (text) at (6.5,-2) {Right rotation};
\end{tikzpicture}

\begin{tikzpicture}
  
  \tikzstyle{noeud}=[circle]
  
   \node[noeud] (z) at (0,0) {$z$};
   \node[noeud] (T4) at (0.5,-1) {$y$};
   \node[noeud] (y) at (-0.5,-1) {$T_1$};
    \node[noeud] (T3) at (1,-2) {$T_4$};

   \node[noeud] (x) at (0,-2) {$x$};
   \node[noeud] (T1) at (-0.5,-3) {$T_2$};
    \node[noeud] (T2) at (0.5,-3) {$T_3$};

   \draw (z)--(y);
   \draw (z)--(T4);
   \draw (x)--(T4);
   \draw (T3)--(T4);
   \draw (x)--(T1);
   \draw (x)--(T2);

   \node[noeud] (z') at (5,0) {$z$};
   \node[noeud] (x') at (5.5,-1) {$x$};
   \node[noeud] (y') at (4.5,-1) {$T_1$};
    \node[noeud] (T4') at (6,-2) {$y$};

   \node[noeud] (T1') at (5,-2) {$T_2$};
   \node[noeud] (T2') at (5.5,-3) {$T_3$};
    \node[noeud] (T3') at (6.5,-3) {$T_4$};

   \draw (z')--(y');
   \draw (z')--(x');
   \draw (x')--(T4');
   \draw (T3')--(T4');
   \draw (x')--(T1');
   \draw (T4')--(T2');
   
   \draw[->] (2,-1.5)--(4,-1.5); 
   \node[noeud] (text) at (3,-2) {Right rotation};
   
   \node[noeud] (x'') at (10,0) {$x$};
   \node[noeud] (z'') at (11,-1) {$y$};
   \node[noeud] (y'') at (9,-1) {$z$};
   \node[noeud] (T3'') at (10.5,-2) {$T_3$};
   \node[noeud] (T4'') at (11.5,-2) {$T_4$};
    \node[noeud] (T2'') at (9.5,-2) {$T_2$};
    \node[noeud] (T1'') at (8.5,-2) {$T_1$};

   \draw (z'')--(x'');
   \draw (z'')--(T4'');
   \draw (x'')--(y'');
   \draw (T3'')--(z'');
   \draw (y'')--(T1'');
   \draw (y'')--(T2'');
   \draw[->] (6,-1.5)--(8,-1.5); 
   \node[noeud] (text) at (6.5,-2) {};
   
  \end{tikzpicture}
  
  \begin{tikzpicture}
   \tikzstyle{noeud}=[circle]
  
   \node[noeud] (z) at (0,0) {$z$};
   \node[noeud] (y) at (0.5,-1) {$y$};
   \node[noeud] (T1) at (-0.5,-1) {$T_1$};
    \node[noeud] (T4) at (1.5,-3) {$T_4$};

   \node[noeud] (x) at (1,-2) {$x$};
   \node[noeud] (T2) at (0,-2) {$T_2$};
    \node[noeud] (T3) at (0.5,-3) {$T_3$};

   \draw (z)--(y);
   \draw (z)--(T1);
   \draw (x)--(T4);
   \draw (T3)--(x);
   \draw (x)--(y);
   \draw (y)--(T2);
   
   \node[noeud] (y') at (7,0) {$y$};
   \node[noeud] (z') at (8,-1) {$x$};
   \node[noeud] (T4') at (8.5,-2) {$T_4$};
   
    \node[noeud] (T3') at (7.5,-2) {$T_3$};

   \node[noeud] (x') at (6,-1) {$z$};
   \node[noeud] (T1') at (5.5,-2) {$T_1$};
    \node[noeud] (T2') at (6.5,-2) {$T_2$};

   \draw (z')--(y');
   \draw (z')--(T4');
   \draw (x')--(y');
   \draw (T3')--(z');
   \draw (x')--(T1');
   \draw (x')--(T2');
   
   \draw[->] (2,-1.5)--(4,-1.5); 
   \node[noeud] (text) at (3,-2) {Left rotation};
  \end{tikzpicture}

\end{figure}

  Assume that $T$ is balanced, and assumed that we now insert $w$ in $T$ following algorithm 
  \ref{Insertionalgo}. Assume further that, after this insertion, $T$ has become unbalanced. 
  We can re-balance $T$ by performing rotations on certain nodes of $T$.
  Starting from $w$, move up towards the root of $T$. We denote by $z$ the first unbalanced node which we encounter on this path. Let 
  $y$ be the heigher child of $z$ and $x$ the heigher child of $y$. There are four 
  possible configurations for the relative positions of $x,y$ and $z$, as shown by Figure \ref{rebalancefigure}. To each configuration, we associate a sequence of rotations, shown in Figure \ref{rebalancefigure}.
  These rotations transform $T_z$ into a balanced tree. 
  It is easy to show that the new tree obtained from $T_z$ by these rotations has the same height as $T_z$ had before inserting $w$. Since $T$ was balanced before inserting $w$, it follows that $T$ is again balanced once the rotations have been performed.
  
Here is a modified version of algorithm \ref{Insertionalgo} which 
gives the insertion algorithm for AVL trees:\\

\begin{algorithm}[H]\label{balancedInsertionalgo}
\SetAlgoLined
\SetKwInOut{Input}{input}\SetKwInOut{Output}{output}

 \Input{Integer $n$, tree $T$}
 Set $node=root(T)$;
 
  \While{$true$}{
  
  \uIf{$n<label(node)$}{
  \uIf{$left\_child(node)=NULL$}{
    $left\_child(node):=create\_node()$;
    
    $label(left\_child(node)):=n$;
    }
 \uElse{
    $node:=left\_child(node)$;}
 }
   
   \uElseIf{$n>label(node)$}{
  \uIf{$right\_child(node)=NULL$}{
    $right\_child(node):=create\_node()$;
    
    $label(right\_child(node)):=n$;
    }
 \uElse{
    $node:=right\_child(node)$;}
 }
  \uElse{
 Exit program;}}
 
 $height(node):=1+max(height(left\_child(node)),heigt(right\_child(node)))$
 
 $balance:=height(left\_child(node))-height(right\_child(node))$
 
  \While{$-1\leq balance\leq 1$}{
  \uIf{$node=root$}{
    return;}
    
    $node:=parent(node)$;
    
    $height(node):=1+max(height(left\_child(node)),heigt(right\_child(node)))$;
    
    $balance=height(left\_child(node))-height(right\_child(node))$;

    }
 \uIf{$balance>1$}{
 \uIf{$n<label(left\_child(node))$}{
    $node=right\_rotate(node)$;}
\uElse{
    $left\_child(node)=left\_rotate(left\_child(node))$;
    
    $node=right\_rotate(node)$;
    }
    }
    \uIf{$balance<1$}{
 \uIf{$n>label(right\_child(node))$}{
    $node=left\_rotate(node)$;}
\uElse{
    $right\_child(node)=right\_rotate(right\_child(node))$;
    
    $node=left\_rotate(node)$;
    }
    }
 \caption{AVL insertion}
\end{algorithm}

Now let us compute the running time of this algorithm. 
The algorithm starts with the naive insertion, which runs in time $O(\log_2(N))$.  
We then have to move back up the tree to update the heights of each node until we find the first unbalanced node. In the worst case, this is done in time 
$O(\log_2(N))$. Finally, the rotation operations are done in $O(1)$ time. Therefore, this algorithm runs in time $O(\log_2(N))$. 
This shows that AVL trees allow for fast look-up and insertion; more precisely, these operations are performed in logarithmic time.

\section{Implementation of ant routing}\label{Implementationsection}

  We saw earlier that the main task of a node is to manage  a set of seeds. The node must 
  frequently operate a look-up on its set of stored seeds and sometimes insert a new seed. 
  It is therefore important to use a data structure allowing fast look-up and insertion. 
  For this reason, we chose AVL trees as the base structure for storing seeds.
  
  A node must store three types of seeds: pheromone, matched and confirmed. Each of these 
  is stored in a separate tree. 
  
  We focus on how pheromone seeds are stored and managed.
  Let $\eta$ be the life time of seeds (in seconds), and set $k:=\frac{\eta}{0.1}$.   
  Pheromone seeds are stored in a tree $T$. The root of $T$ has $k+1$-many children. Each of these children is the root 
  of an AVL tree. We denote by $T_0,T_1,\dots,T_k$ these subtrees of $T$.
  Each $T_i$ corresponds to a time interval
  $I_i:=[t_0+i*0.1,t_0+(i+1)*0.1[$. A seed with timestamp $t$ is stored in the tree $T_i$ such that 
  $t\in I_i$. If the node receives a seed with a timestamp outside of the time interval 
  $[t_0,t_0+(k+1)*0.1[$, then the seed is discarded.
  At time $t_1:=t_0+(k+1)*0.1$, the node erases the tree $T_0$, which contains seeds older than $\eta$, and 
  creates a new tree $T_{k+1}$. This new tree will contain the seeds with timestamp 
  in $[t_1,t_1+0.1[$. At time $t_1+0.1$, the node will erase $T_1$, create $T_{k+2}$, and the process continues.
  
  Each $T_i$ is an AVL tree labeled by seeds.
  Each node of the tree
  contains the following fields: \\
  
  \begin{itemize}
   \item An integer $seed$. This is the random number $S$ chosen by Alice and Bob. This also serves as the label of the node.
   \item An integer $amount$, indicating the amount of the transaction.
   \item A boolean $pheromone0$. The value is true if and only if $P(0)$ has been received.
   \item A boolean $pheromone1$. The value is true if and only if $P(1)$ has been received.
   \item An integer $fees0$, giving the remaining fees associated to $P(0)$ if $P(0)$ was received, and set to $0$ otherwise.
   \item An integer $fees1$, giving the remaining fees associated to $P(1)$ if $P(1)$ was received, and set to $0$ otherwise.
   \item An integer $counter0$. This is the counter associated to $P(0)$, set to $0$ if $P(0)$ was never received.
   \item An integer $counter1$. This is the counter associate to $P(1)$, set to $0$ if $P(1)$ was never received.
   \item An integer $sender0$. This is the node which sent  $P(0)$, set to $0$ if $P(0)$ was never received.
   \item An integer $sender1$. This is the node which sent $P(1)$, set to $0$ if $P(1)$ was never received.
   \item A pointer $left\_child$.
   \item A pointer $right\_child$.
  \end{itemize}

  This describes the pheromone tree. The match tree is built in a similar way. The only difference between 
  pheromone and match trees resides in the structure of the nodes. In the match tree, each tree 
  $T_i$ is labeled with a match identifier $Id$. Each node of $T_i$ contains the following fields: 
  
  \begin{itemize}
   \item An integer $Id$ (match identifier).
   \item An integer $target$.
   \item A pointer $left\_child$.
   \item A pointer $right\_child$.   
  \end{itemize}

  The confirmation tree is  similar to the pheromone and the match tree, except for the structure of nodes.
  Each node of the confirmation tree contains the following fields: 
  
  \begin{itemize}
   \item An integer $Id$ (match identifier).
   \item An integer $target$.
   \item An integer $check$.
   \item A pointer $left\_child$.
   \item A pointer $right\_child$.   
  \end{itemize}
  
  We now give an estimate of the memory space used by the seeds with this implementation.
  The following table gives the size in Bytes of a node of each tree: \\
  \tabcolsep=1pt\relax
  \begin{center}
  \begin{tabular}{|c|c|c|c|c|c|c|c|c|}
  \hline
   & seed  & counter & amount & fees & sender/target & children & check & total\\
   \hline
   pheromone & $8$ & $1$ & $4$ & $4$ & $1$ & $16$ & $0$ & $34$\\
   \hline 
   match & $8$ & $0$ & $0$ &  $0$ & $1$ & $16$ & $0$ & $25$\\
   \hline
   confirmation & $8$ & $0$ & $0$ & $0$ & $1$ & $16$ & $8$ & $33$\\
   \hline
  \end{tabular}
  \end{center}
  \vspace{10pt}
  Note also that biggest nodes have around 100 neighbors, which is why $1$ Byte is enough to store the sender or target. 
  Pointers typically have size $8$ Bytes, which is why a tree node needs $16$ Bytes to store the addresses of its children. 
     We can check if the length of $8$ Bytes for seeds is sufficient to avoid collisions. We can estimate the probability of collisions occuring by using an approximation formula of the birthday problem. If $n$ is the number of seeds simultaneously present in the network and 
 $N$ the number of all possible seeds, then the probability $p$ of having a collision at a given time is 
 $p\simeq\frac{n^2}{2N}$. We have $n=\lambda\eta$, where $\lambda$ is the rate of incoming transactions and $\eta$ the life time of seeds. If we take $\lambda=10000$ and $\eta=2$, then we get 
 $n=20000$. For seeds of length $8$ Bytes we have $N=2^{64}$. Therefore, the probability of having a collision at a given time is 
 $p\simeq\frac{(20000)^2}{2^{65}}$. The probability of having a collision in $100$ years is then 
  $1-(1-\frac{(20000)^2}{2^{65}})^{3600\times 24\times 31\times 12\times 100}$, which is approximately $3$\%.
 
 If one wishes to improve these odds, one can increase the length of seeds to 9 Bytes. In that case, the probability of 
 having a collision in $100$ years is $1-(1-\frac{(20000)^2}{2^{82}})^{3600\times 24\times 31\times 12\times 100}$, which is approximately 
 $2.10^{-5}$.
 
 Note that the simulations whose results we present in Section \ref{experimentalsection} were done with seeds of size $8$ Bytes. 
 However, increasing the length of seeds to $9$ Bytes will have a negligeable effect on the computation time needed for the 
 treatment of seeds.

  \vspace{10pt}

  Now let us compute the total memory $M$ taken in one node for the routing task. 
  For one routing task, a node stores at most one pheromone seed and one confirmed seed, but potentially several matched seeds. 
  Therefore, The memory taken in a node for one routing task is upper-bounded by 
  $34+25r+33$ Bytes, where $r$ is the number of matched received. Let $\lambda$ be the rate of transactions, i.e. the average number 
  of transactions performed on the network at each second. Then the number of seeds kept in memory by a node is at most 
  $\lambda\eta$, where $\eta$ is the life time of seeds, so the maximal amount of memory taken is 
  $\lambda\eta(34+25r+33))$.
  
  We argued above that we can expect to set $\eta:=2$.  Setting $\lambda=10000$, we get a 
  $M=20000\times (67+25r)$. For smaller, less connected nodes, $r$ should take small values. 
  For $r\leq 8$,  we get $2$MB$\leq M\leq 4$MB. 
  For bigger nodes, $r$ may take bigger values, in which case we get (for $r> 8$)
  $M\simeq 500000r=0.5r$MB.

  Finally, we want to estimate the necessary bandwidth for the propagation of pheromones. 
  Remember that pheromone seeds are relayed as messages of the form 
  $(P,c,f,a,t)$ (see Section \ref{memorysubsection}).  The size in Bytes of one such message is given by the following table:\\ 
  
  \begin{center}
  \begin{tabular}{|c|c|c|c|c|c|}
  \hline
    pheromone seed  & counter & fees & amount & timestamp & total\\
   \hline
    $8$ & $1$ & $4$ & $4$ & $1$ & $16$\\
   \hline 
  \end{tabular}
  \end{center}
  \vspace{10pt}

   With a transaction rate of $\lambda=10000$, this means we need a bandwidth of
  $16\times 10000$ Bytes per second, i.e $160$kB/s. 
   The parameter amount above is the amount of the transaction. By comparison with the Bitcoin network, $16$ Bytes is half the size of the hash of a Bitcoin transaction that is sent to nodes in an ``inventory message'' to announce a new transaction. See \cite{Erlay,TxProbe} for details on bitcoin transactions propagation.

  \begin{Rem}
   Note that a standard timestamp like the unix timestamp is 4 Bytes long. However, we propose here to use timestamps of length 1 Byte. Here is how we think timestamps should work in ant routing: each timestamp indicates a time interval of 0.1 seconds 
  \textit{modulo 20 seconds}. Since there are only 200 such time intervals within 20 seconds, 1 Byte is enough to express every possible timestamp. 
  We argue that these small timestamps are enough to make ant routing function. Because seeds are stored in trees each corresponding to a time interval of 0.1 seconds, it is indeed enough to have timestamps with precision 0.1 second. Moreover, we argued in section \ref{lifetimesection} 
  that the life time of seeds can be chosen to be 2 seconds. This means that every seed can be removed after staying more than 2 seconds in memory. 
  It follows that couting timestamps modulo 20 seconds is engouh to manage seeds.
  
  \end{Rem}

\section{Estimating scalability}\label{experimentalsection}

    We address now the following question: is the ant routing protocol compatible with the goal of 
    solving bitcoin's scalability problem? In other words, how many routing requests per second can the network process using ant routing? 
    In view of the arguments given in section \ref{proxymodelsection}, we answer this 
    question by computing the time taken by a node to process all incoming routing requests. 
    This is given by formula (\ref{secondformula}) below, which depends on the rate $\lambda$ of transactions. 
    In section \ref{numericalvaluessection}, we give the numerical values for the parameters appearing in formula (\ref{secondformula}), which we obtained experimentally. This allows us to give a numerical estimate of the maximal transaction rate which the network can sustain.

   \subsection{Formula}
   We now want to estimate the time taken by a node to perform the routing task. 
   We showed in Section \ref{AVLsection} that 
   look-up and insertion in an AVL tree both run in logarithmic time. The operation of deleting a tree is linear in $N$. We 
   respectively denote by $\alpha$,
   $\beta$ and $\gamma$ the numbers such that, for an AVL tree $T$ with $N$-many nodes, look-up in $T$ runs in $\alpha\log_2(N)$ seconds,
   Insertion in $T$ runs in $\beta\log_2(N)$ seconds, and deleting the whole tree takes $\gamma N$ seconds.

   Let us estimate the time taken by a node to process all pheromone seeds for one routing task. 
   We denote by $\lambda$ the rate of transactions, i.e. the number of transactions performed per second.
   In our implementation of ant routing given in Section \ref{Implementationsection}, the tree containing pheromone seeds is split into 
   $k+1$-many subtrees. We can assume that the routing tasks arrive at a constant rate, so that each 
   subtree has the same number of nodes. In that case, each subtree has
   $\frac{\lambda}{10}$-many nodes.     
   The node will only have to perform one insertion of pheromone seed, but may have to perform several look-ups. Let $p$ be the number of look-ups 
   to be performed.  
   Then the total time taken to process the pheromone seeds of one routing task is 
   $p\alpha\log_2(\frac{\lambda}{10})+\beta\log_2(\frac{\lambda}{10})$.
   
   Now let us consider the match phase. 
   Now the trees containing the matches are labelled by the identifier $Id$ of the match. 
   In general, a node may receive several matches produced by the same pheromone seed.  Denote by $m$ the average 
   number of matches created by one pheromone seed and received by a node.
   The node will need to perform $m$-many insertions (one for each match received) on the match tree, but only 
   one look-up (the look-up will be performed during the confirmation phase).  
   The total time taken to process matched seeds is on average 
   $\alpha\log_2(m\frac{\lambda}{10})+m\beta\log_2(m\frac{\lambda}{10})$.
   
   Finally, the confirmation seed only needs one insertion and one look-up (the look-up is actually performed during the 
   counter check phase). 
   Note that, for a given transaction, only very few nodes will actually have to process a confirmation seed. 
   Let $c$ denote the probability that a node receives a confirmation seed.
   The time to process confirmed seeds is then on average  
   $c(\alpha\log_2(c\frac{\lambda}{10})+\beta\log_2(c\frac{\lambda}{10}))$. 
   
   The total time to process one routing task is therefore on average: 
   
   \begin{equation}\label{formula}
   T(\lambda):=(p\alpha+\beta)\log_2(\frac{\lambda}{10})+(\alpha+m\beta)\log_2(m\frac{\lambda}{10})+c(\alpha+\beta)\log_2(c\frac{\lambda}{10})
   \end{equation}
   Now let us  estimate the total time that a node must dedicate to the routing task 
   to process all the routing demands that arise in the span of one second.  The node has to process on average 
   $\lambda$-many routing tasks, so the time taken to process seeds is 
   $\lambda T(\lambda)$. We must also take into account the time taken to clean old seeds. This is done by deleting 
   a tree every $0.1$ seconds. The time dedicated to 
   cleaning the mempool is therefore 
   $\gamma\lambda(1+m+c)$. Thus, the total time dedicated to the ant routing algorithm is 
   
   \begin{equation}\label{secondformula}
   \widehat{T}(\lambda):=\lambda (T(\lambda)+\gamma(1+m+c))
   \end{equation}
   
   Now we just need to estimate the value of the coefficients $\alpha,\beta,\gamma,p,m,c$.
   
   \subsection{Experimental result}\label{numericalvaluessection}
   
   We do determine $\alpha$, $\beta$ and $\gamma$ experimentally.
   We have implemented the work of a node with a C program following Section \ref{Implementationsection}.     
    Our program treats seeds stored in an AVL tree. 
We use a look-up and insertion algorithm in our program that is a recursive Implementation of algorithms \ref{lookupalgo} and \ref{balancedInsertionalgo}.
Our program first randomly generates a mempool of a given size $N$.  
The program then inserts a new randomly generated seed in the tree and records the time it takes 
to the program to perform this insertion.

We ran the program for different values of $N$ in the range from $100$ to $100000$. For each $N$, we ran a Montecarlo simulation 
executing the program with $1000$  trials. We also did the same for the tasks of look-up and cleaning of the mempool. 
The experimental results are shown in Figure \ref{graphs} below. This allowed us to estimate the values of $\alpha,\beta$ and $\gamma$ 
appearing in formula (\ref{secondformula}). Experimentally, we find $\alpha=0.7\times 10^{-6}$, $\beta=1.1\times 10^{-6}$ and $\gamma=8.2\times 10^{-8}$.

\begin{figure}[H]
 \includegraphics[scale=0.18]{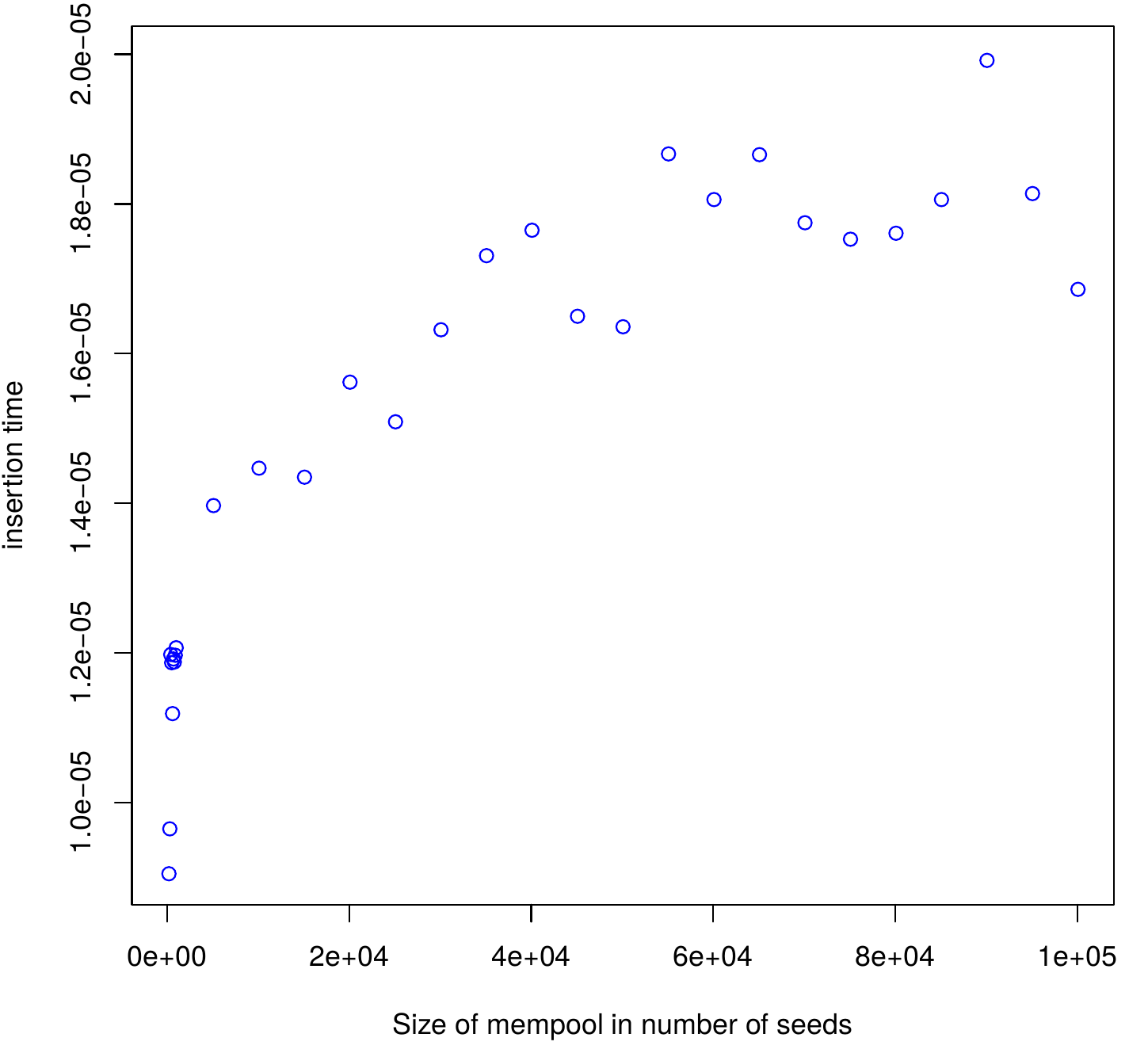}
 \includegraphics[scale=0.18]{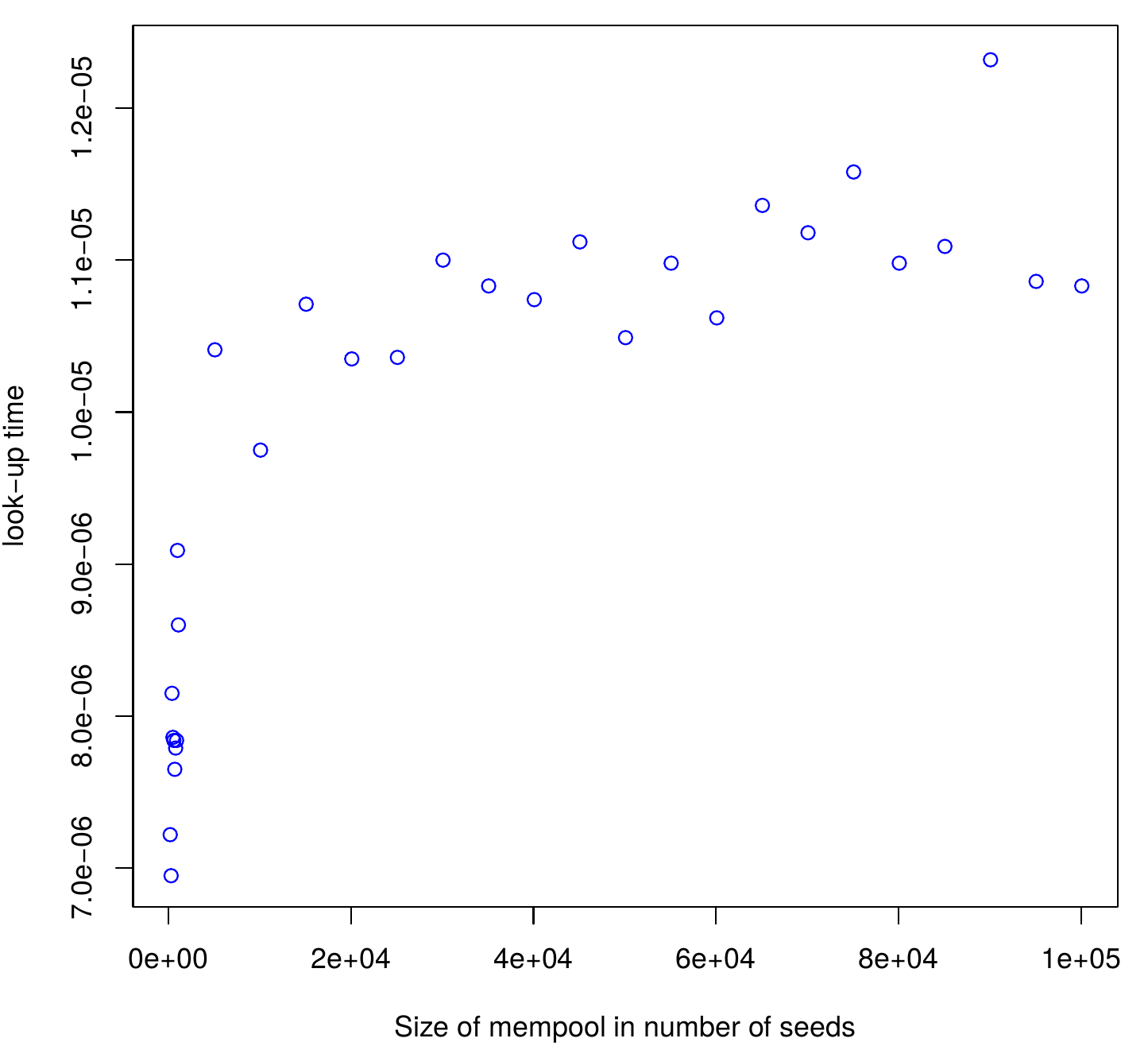}
 \centerline{\includegraphics[scale=0.18]{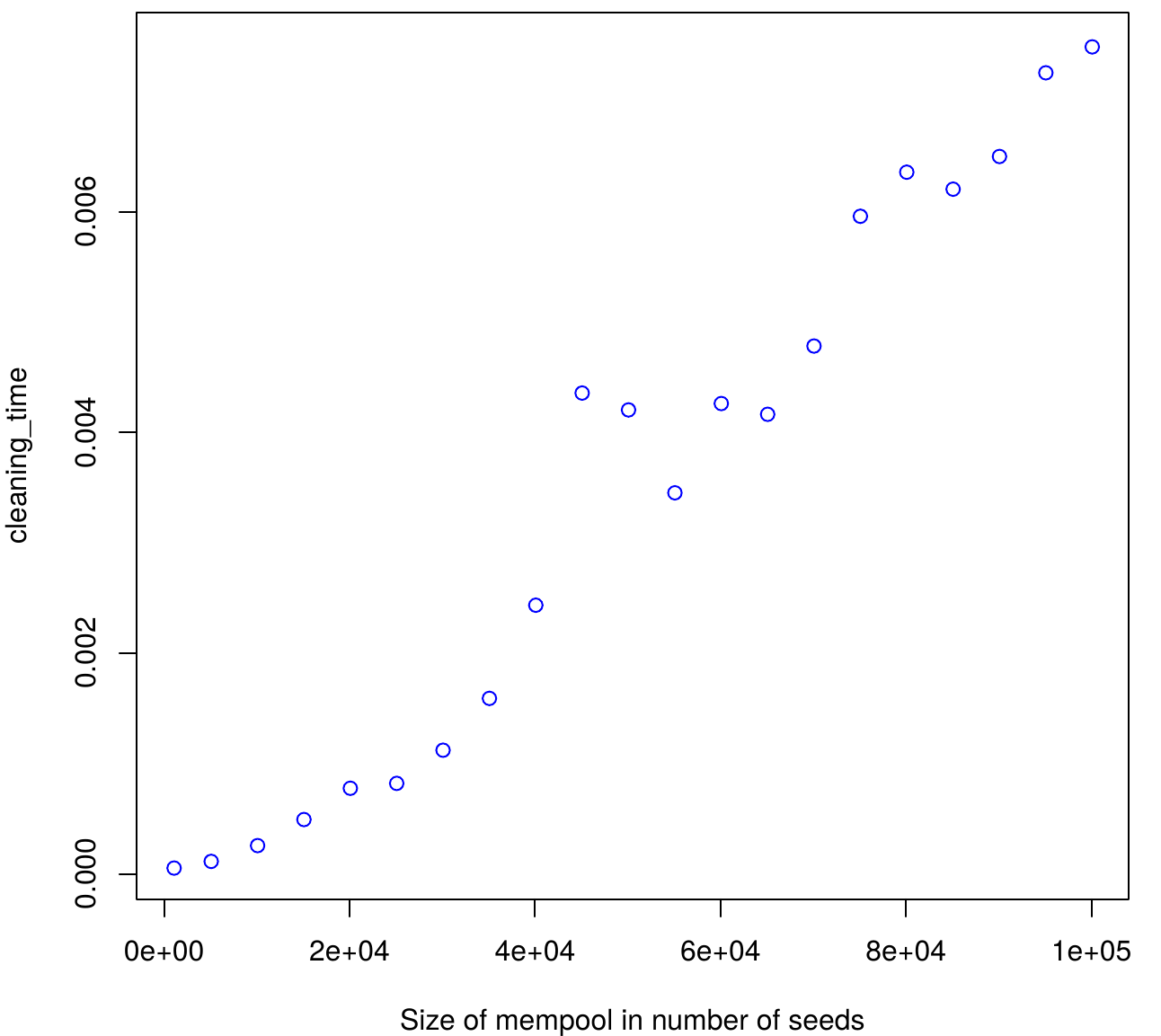}}
 \caption{Insertion, look-up and cleaning time depending on the size of the mempool}\label{graphs}
\end{figure}

Now we come back to formula (\ref{secondformula}).
We can now determine the maximum number of routing tasks which a node can perform per second. In other words,  the maximum 
$\lambda$ such that 
$\widehat{T}(\lambda)<1$. The value of this maximal $\lambda$ depends on parameters 
$p,m$ and $c$. Note that these parameters essentially depend on the centrality and connectivity of the node. 

For most nodes, we can assume that  $c$ is negligible, and that $m=1$ on average, so it remains to determine $p$.
It is established in \cite{LNtopology} that the average number of channels per node is $7$. This means that $p$ is bounded by $8$. In the worst 
case, we thus have 
$T(\lambda)=(8\alpha+2\beta)\log_2(\frac{\lambda}{10})=7.8\log_2(\frac{\lambda}{10})10^{-6}$ and 
$\widehat{T}(\lambda)=\lambda(7.8\log_2(\frac{\lambda}{10})10^{-6}+8.2\times 10^{-8}\times 2)$.
 We find 
$\lambda_{\max}\simeq 12500$.

\begin{Rem}\label{endrem}
\begin{itemize} 
 \item This result means that an average node can process up $12500$ tx per second. However, it is important to note that it is not necessary that 
 all nodes process all incoming routing tasks. Indeed, since the network is well-connected, a routing task should still 
 be completed successfully even if only 50 percent of the network is  performing the task. This means that we can hope 
 that the network could sustain up to $25000$ tx per second.
 \item  The above estimate is valid for nodes with low or average connectivity. However, for big nodes, $p$ can be much higher, the value of 
 $c$ might be non-negligible, and $m$ might take an average value higher than $1$. It would take a deeper analysis of the network 
 to estimate the capacity of these nodes to process all incoming routing tasks.

\end{itemize}

\end{Rem}

The C code used in this section for estimating the capacity of Ant Routing algorithm is available at :
https://github.com/gabrielLehericy/Ant-routing-simulation
 
\medskip

\textbf{Acknowledgement} The two first authors were partially supported by the FUI Moneytrack project joined between INRIA and Pôle Universitaire Léonard de Vinci.

\end{document}